\documentclass{aa}

\usepackage{graphicx}
\usepackage{txfonts}
\usepackage{natbib}
\usepackage{booktabs}
\usepackage{multirow}
\usepackage{longtable}
\usepackage{url}
\usepackage[rightcaption]{sidecap}

\begin{document}

   \title{Quiet-Sun hydrogen Lyman-$\alpha$ line profile derived from SOHO/SUMER solar-disk observations}

   \author{S. Gun\' ar\inst{1} and P. Schwartz\inst{2} and J. Koza\inst{2} and P. Heinzel\inst{1}}

   \institute{
   Astronomical Institute,
        The Czech Academy of Sciences,
        25165 Ond\v rejov, Czech Republic\\
        \email{stanislav.gunar@asu.cas.cz}
         \and
         Astronomical Institute of Slovak Academy of Sciences, 05960 Tatransk\'a Lomnica, Slovak Republic\\
         }

   \abstract
   {The solar radiation in the Lyman-$\alpha$ spectral line of hydrogen plays a significant role in the illumination
of chromospheric and coronal structures, such as prominences,
spicules, chromospheric fibrils, cores of coronal mass ejections,
    and solar wind. Moreover, it is important for the investigation
    of the
    heliosphere, Earth's ionosphere, and the atmospheres of planets,
    moons, and comets.}
   {We derive a reference quiet-Sun Lyman-$\alpha$ spectral profile
   that is representative of the Lyman-$\alpha$ radiation from the
   solar disk during a minimum of solar activity. This profile can
    serve as an incident radiation boundary condition for the
    radiative transfer modelling of chromospheric and coronal
    structures. Because the solar radiation in the Lyman lines is
    not constant over time but varies significantly with the solar
    cycle, we provide a method for the adaptation of the incident     radiation Lyman line profiles (Lyman-$\alpha$ and higher
    lines) to a specific date. Moreover, we analyse how
    the change in the incident radiation influences the synthetic spectra produced by
    the radiative transfer modelling.}
   {We used SOHO/SUMER Lyman-$\alpha$ raster scans obtained without
   the use of the attenuator in various quiet-Sun regions on
   the solar disk. The observations were performed on three
   consecutive days (June 24, 25, and 26, 2008) during a period of
   minimum solar activity. The reference Lyman-$\alpha$ profile
   was obtained as a spatial average over eight available raster
   scans. To take into account the Lyman-$\alpha$
   variation with the solar
   cycle, we used the LISIRD composite Lyman-$\alpha$ index. To estimate the influence of the
   change in the incident radiation in the Lyman lines on the
   results of radiative transfer models, we used a 2D prominence fine structure model.}
   {We present the reference quiet-Sun Lyman-$\alpha$ profile
   and a table of coefficients describing the variation of the Lyman lines with the solar cycle throughout the lifetime of SOHO. The
   analysis of the influence of the change in the incident radiation shows that the
   synthetic spectra are strongly affected by the modification of the incident radiation boundary condition. The most pronounced
   impact is on the central and integrated intensities of
   the Lyman lines. There, the change in the synthetic spectra can
   often have the same amplitude as the change in the incident
   radiation itself. The impact on the specific intensities in the
   peaks of reversed Lyman-line profiles is smaller but still
   significant. The hydrogen H$\alpha$ line can also be considerably affected, despite the fact that the H$\alpha$ radiation from the solar disk does not vary with the
   solar cycle.}
   {}

   \authorrunning{Gun\' ar, Schwartz, Koza \& Heinzel 2020}
   \titlerunning{Quiet-Sun hydrogen Lyman-$\alpha$ line profile}
   \maketitle
%

\section{Introduction}

The Lyman-$\alpha$ line of hydrogen (1215.67\,$\AA$) is the most
intense ultraviolet (UV) spectral line observed on the solar disk.
As such, it plays a significant role in the illumination of
chromospheric and coronal structures, such as prominences,
spicules, chromospheric fibrils, solar wind, and cores of coronal
mass ejections (CMEs). A well-determined Lyman-$\alpha$ spectrum
of the solar disk is thus one of the crucial boundary conditions
for the radiative transfer models of these structures. The
Lyman-$\alpha$ illumination is often referred to as the
Lyman-$\alpha$ incident radiation in the modelling context.
Details on the radiative transfer modelling of the photosphere,
chromosphere, and corona, are available in, for example, the
review by \citet{2020LRSP...17....3L}. Prominence modelling was
reviewed by, for example, \citet{2018LRSP...15....7G} and
\citet{2014IAUS..300...59G}. An example of a recent study
employing radiative transfer modelling in spicules is available
in, for example, \citet{2020ApJ...888...42T}, and an example of
filament modelling is available in \citet{2019A&A...631A.146S}.
Information regarding the modelling of Lyman-$\alpha$ in CME cores
is available in \citet{2016A&A...589A.128H}, and information
regarding the modelling of Lyman-$\alpha$ in the solar wind is
available in \citet{2019A&A...627A..18D}. \vspace{0.2 cm}

Both the solar Lyman-$\alpha$ line profile and the total
Lyman-$\alpha$ irradiance (i.e. the radiant flux received at a
surface per unit area integrated over the wavelength range of
Lyman-$\alpha$) are also important for the investigation of the
Earth's ionosphere \citep[see
e.g.][]{2013JGRA..118..570R,2014AdSpR..54.1276N}, the heliosphere
\citep[see
e.g.][]{2013ccfu.book...67B,2013ccfu.book..177G,2017A&A...598A..12K},
cometary atmospheres \citep[see
e.g.][]{2013ccfu.book..255P,2019Icar..317..610C}, the Moon
\citep[see e.g.][]{2020LPI....51.1665P,2020LPI....51.2781R}, Mars
\citep[see e.g.][]{2017JGRA..122.1296B,2017JGRA..122.2748T}, and
other planets. \vspace{0.2 cm}

While the total Lyman-$\alpha$ irradiance is continuously
monitored by a succession of instruments -- SOLSTICE/UARS
\citep[][]{1994SPIE.2266..317R}, XPS/SORCE
\citep[][]{2005SoPh..230..345W}, and LYRA/PROBA2
\citep[][]{2013SoPh..286...21D} -- precise spectroscopic
observations of the Lyman-$\alpha$ line on the solar disk are
rare. This is because the Lyman-$\alpha$ line lies in the UV part
of the spectrum, and, as such, it needs to be observed from above
the Earth's atmosphere -- and preferably above the geo-corona,
which absorbs in the Lyman-$\alpha$ core. Moreover, its very high
intensity and extreme UV wavelength range complicate the
space-borne spectroscopic observations. Previously, calibrated
solar disk Lyman-$\alpha$ (and Lyman-$\beta$) profiles obtained by
the LPSP (Laboratoire de Physique Stellaire et Plan\'etaire,
\citealt{1977SSI.....3..131A} and \citealt{1978ApJ...221.1032B})
on board the OSO-8 (Orbiting Solar Observatory) were presented in
\citet{1978ApJ...223L..55L} and \citet{1978ApJ...225..655G}. Solar
disk Lyman-$\alpha$ observations were also obtained by
\citet{1976SoPh...46...53R} using Skylab ATM (Apollo Telescope
Mount) and by the HRTS (High-Resolution Telescope and
Spectrograph) rocket-flight instrument
\citep[][]{1979ApJ...230..924B}. Later, Lyman-$\alpha$ spectra of
on-disk structures obtained using the Ultraviolet Spectrometer and
Polarimeter (UVSP) instrument on board the SMM (Solar Maximum
Mission) were analysed by \citet{1988ApJ...329..464F}.
High-resolution spectro-polarimetric Lyman-$\alpha$ observations
of the solar disk were recently obtained by
\citet{2017ApJ...839L..10K} using the Chromospheric Lyman-Alpha
Spectro-Polarimeter (CLASP) rocket experiment. All these
observations were made within the geo-corona and were thus
affected by absorption in the Lyman-$\alpha$ line core. In the era
of the SOHO (Solar and Heliospheric Observatory), which is located
at the Lagrangian point L1 outside the influence of geo-corona,
solar disk Lyman-$\alpha$ spectra were deduced by
\citet{1998A&A...334.1095L} from the scattered light at the
primary mirror of the SOHO/SUMER \citep[Solar Ultraviolet
Measurements of Emitted Radiation,][]{1995SoPh..162..189W}
spectrograph. SOHO/SUMER typically observed solar-disk
Lyman-$\alpha$ at a dedicated part of the detectors with an
attenuating layer to protect them from the deterioration caused by
the high Lyman-$\alpha$ intensity. This attenuator, however, was
introduced non-linearity into the radiometric calibration, which
makes the calibration of Lyman-$\alpha$ spectra difficult and
unreliable. Only a small number of SOHO/SUMER Lyman-$\alpha$
observations without the use of the attenuator were carried out.
The first such Lyman-$\alpha$ observations (obtained on June 24
and 25, 2008) with reduced incoming photon flux due to the
partially closed aperture door of the telescope were presented and
analysed by \citet{2008A&A...492L...9C}. Later,
\citet{2009A&A...504..239T,2009ApJ...703L.152T} analysed further
SOHO/SUMER Lyman-$\alpha$ observations of the quiet Sun and polar
coronal holes. In the current work, we used the quiet-Sun
Lyman-$\alpha$ observations that were obtained over three
consecutive days between June 24 and 26, 2008. This set of eight
SOHO/SUMER raster scans allows us to derive the reference
quiet-Sun Lyman-$\alpha$ spectra unaffected by the geo-corona. In
doing so, we complement the comprehensive study of the Lyman line
series above the Lyman-$\alpha$ line carried out by
\citet{1998ApJS..119..105W}.\vspace{0.2 cm}
\begin{figure*}
\centerline{\includegraphics[width=18cm]{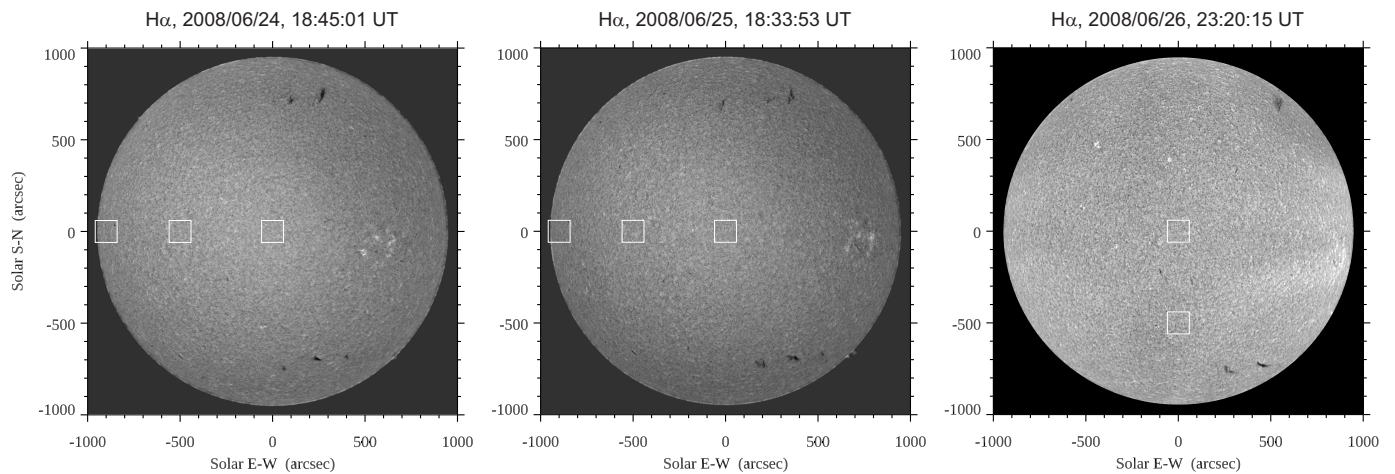}}
\caption{Context H$\alpha$ images obtained by the Big Bear Solar
Observatory on June 24, 2008, at 18:45:01\,UT (left) and on June
25, 2008, at 18:33:53\,UT (middle), and by the SMART telescope at
the Hida Observatory (right) on June 26, 2008, at 23:20:15\,UT.
Squares indicate the positions of the SOHO/SUMER rasters used
here. The contrast of the H$\alpha$ images was enhanced, and the
limb darkening was corrected.} \label{fig:context}
\end{figure*}

The Lyman-$\alpha$ radiation from the solar disk is not constant
over time but varies significantly with the solar cycle. This
variation is well demonstrated by, for example, the LISIRD
composite Lyman-$\alpha$
index\footnote{\url{lasp.colorado.edu/lisird/data/composite_lyman_alpha}}
\citep[][]{2019E&SS....6.2263M}. In extreme cases, the values of
this index exhibit differences between the minima and maxima of a
cycle of up to 100\%. An empirical model of the solar
Lyman-$\alpha$ irradiance variation was developed by
\citet{2018GeoRL..45.2138K}. This model is in a good agreement
with the LISIRD composite Lyman-$\alpha$ index. Variations of the
Lyman-$\alpha$ (and Lyman-$\beta$) spectra over solar cycle 23
were also studied by
\citet{2005AdSpR..35..384L,2015A&A...581A..26L}. These authors
used the technique of \citet{1998A&A...334.1095L} and observed the
scattered light in SOHO/SUMER to derive full-disk Lyman-$\alpha$
profiles throughout cycle 23. In the present study, we complement
their work by using the SOHO/SUMER solar disk Lyman-$\alpha$
observations and the LISIRD composite Lyman-$\alpha$ index
\citep[][]{2019E&SS....6.2263M} to derive the variation of the
Lyman-$\alpha$ spectra throughout cycle 24 and beyond. \vspace{0.2
cm}

In addition, in the present paper, we investigate the influence of
the change in the illumination in the Lyman lines on the results
of radiative transfer models. To do so, we used as an example the
synthetic spectra produced by the 2D non-LTE (i.e. departures
from local thermodynamic equilibrium) prominence vertical fine
structure model from \citet{2001A&A...375.1082H}. These synthetic
spectra, which are based on a realistic set of model input
parameters \citep[see][]{2010A&A...514A..43G}, allowed us to
estimate the degree by which the variation of the illumination
during the solar cycle affects the Lyman lines as well as the
H$\alpha$ line.\vspace{0.2 cm}

The paper is organized as follows. The SOHO/SUMER observations
used for the derivation of the reference full-disk Lyman-$\alpha$
line profile are described in Sect.~\ref{Sect:Obs}. The variation
of the Lyman spectra with the solar cycle is addressed in
Sect.~\ref{Sect:cycle}. The influence of the illumination change
on the synthetic spectra produced by prominence radiative transfer
models is analysed in Sect.~\ref{Sect:influence}. In
Sect.~\ref{Sect:discussion}, we discuss the results, and  we offer
our conclusions in Sect.~\ref{Sect:conclusions}. Tables containing
the reference Lyman-$\alpha$ profile can be found in
Appendix~\ref{App:Lya} and as an electronic attachment. In
Appendix~\ref{App:cycle}, we provide a table of coefficients
describing the variation of the Lyman spectra throughout the
lifetime of SOHO. These coefficients can be also downloaded as an
electronic attachment.
\begin{figure*}
\centerline{\includegraphics[width=16cm]{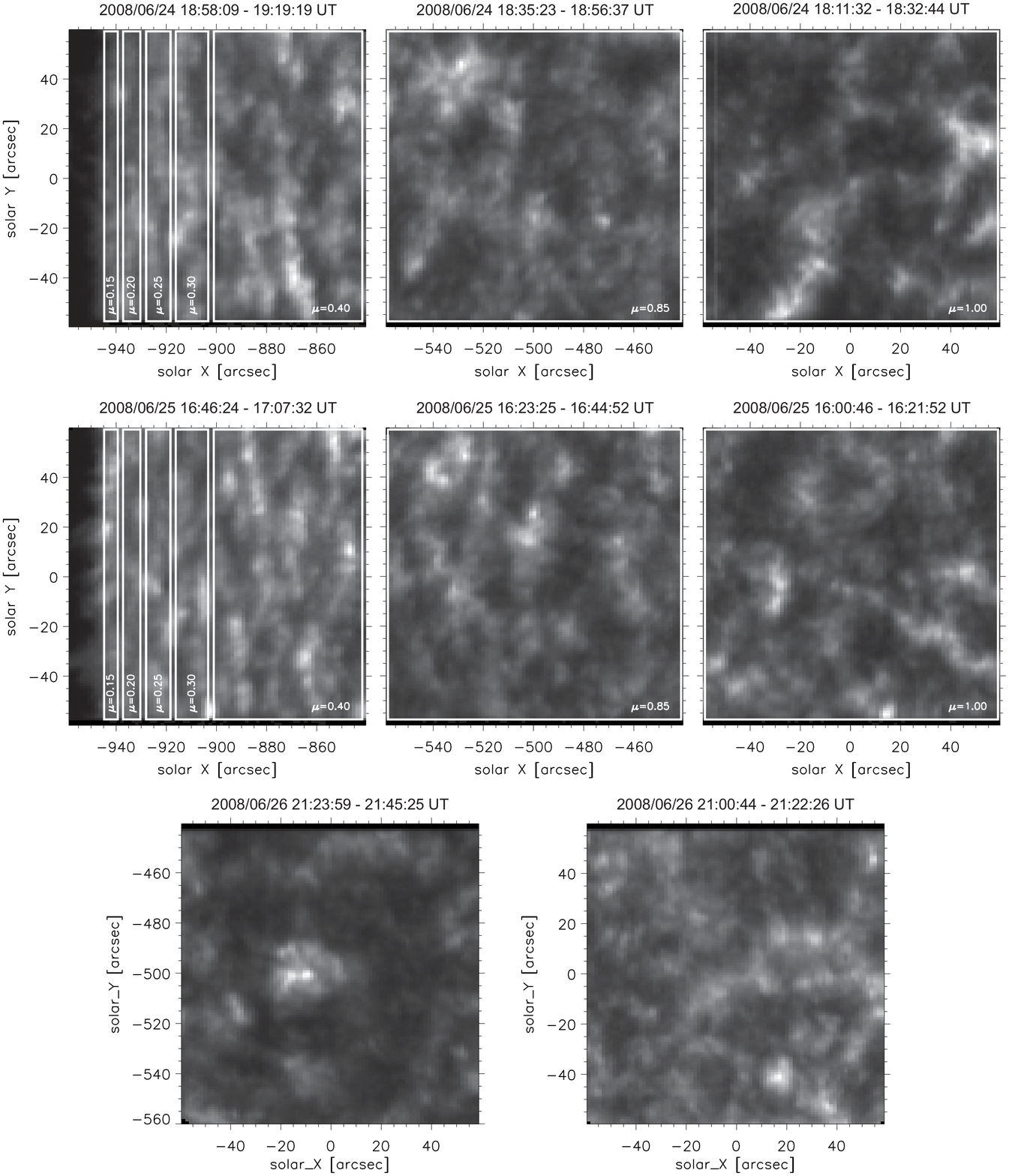}}
\caption{Lyman-$\alpha$ integrated intensity maps in SOHO/SUMER
rasters obtained on June 24, 2008 (top row), June 25, 2008 (middle
row), and June 26, 2008 (bottom row). White rectangles in rasters
in the upper and middle rows highlight the regions selected for
spatial averaging. These regions correspond to
Fig.~\ref{fig:mu+I}. Each region is centred at the marked $\mu$
value.} \label{fig:Rasters}
\end{figure*}


\section{SOHO/SUMER observations}\label{Sect:Obs}

\renewcommand\arraystretch{1.3}
\begin{table*}
  \caption{List of the used SOHO/SUMER Lyman-$\alpha$ raster
  scans. The solar $X$ and $Y$ coordinates give the position of the
  centre of individual rasters.
  }
  \normalsize
  \centerline{
    \begin{tabular}{cccccc}
      \toprule
      Raster position & \,Solar $X$ [arcsec]\, & \,Solar $Y$ [arcsec]\, & \quad\quad $\mu$ interval \quad\quad\, & \quad\quad date \quad\quad\quad & time [UT] \\
      \midrule
      disk centre & 0 & 0 & 1.0 $\pm$0.002 & $2008/06/24$ & 18:11:32 -- 18:32:44\\
      midway & $-500$ & 0 & 0.89 -- 0.81 & $2008/06/24$ & 18:35:23 -- 18:56:37\\
      near limb & $-900$ & 0 & 0.47 -- 0.00 & $2008/06/24$ & 18:58:09 -- 19:19:19\\
      disk centre & 0 & 0 & 1.0 $\pm$0.002 & $2008/06/25$ & 16:00:46 -- 16:21:52\\
      midway & $-500$ & 0 & 0.89 -- 0.81 & $2008/06/25$ & 16:23:25 -- 16:44:52\\
      near limb & $-900$ & 0 & 0.47 -- 0.00 & $2008/06/25$ & 16:46:24 -- 17:07:32\\
      disk centre & 0 & 0 & 1.0 $\pm$0.002 & $2008/06/26$ & 21:00:44 -- 21:22:26\\
      midway & 0 & $-500$ & 0.89 -- 0.81 & $2008/06/26$ & 21:23:59 -- 21:45:25\\
      \toprule
  \end{tabular}}
  \label{tab:Obs}
\end{table*}
\renewcommand\arraystretch{1.0}
We used a set of eight SOHO/SUMER Lyman-$\alpha$ raster scans of
various quiet-Sun regions obtained with reduced incoming photon
flux on June 24, 25, and 26, 2008. These observations correspond to
the minimum of solar cycle 23. On June 24 and 25, the rasters were
pointed at the disk centre ($\mu$=1.0), midway between the disk
centre and the east limb ($\mu$=0.85), as well as near the east limb
($\mu$ from 0.47 to 0.0). On June 26, pointing was at the disk
centre and midway between the disk centre and the south pole
($\mu$=0.85). More details are available in Table~\ref{tab:Obs}. In
Fig.~\ref{fig:context}, we show the position of the rasters on the
full-disk H$\alpha$ images obtained by the Big Bear Solar
Observatory and by the SMART telescope \citep{2004ASPC..325..319U}
at the Hida Observatory.  In Fig.~\ref{fig:Rasters}, we plot the
integrated intensity maps in each raster. All rasters have
dimensions of $120\times120$~arcsec and consist of 80 slit
positions with N-S orientations. The slit has dimensions of
$0.28\times120$~arcsec. The exposure time is 15~sec, and the step
between slit positions is 1.5~arcsec. The spectral data were
recorded on the bare part of detector B (i.e. outside the
attenuator) with the standard voltage of 5656~V. The
Lyman-$\alpha$ spectral profile was obtained separately in two
halves and subsequently stitched together.\vspace{0.2 cm}

All Lyman-$\alpha$ observations used in this work were obtained
with a partially closed aperture door of the telescope to decrease
the incoming photon flux and thus protect the SOHO/SUMER detector.
A detailed description of this observational technique is given in
\citet{2008A&A...492L...9C}. The observed data were reduced and
calibrated using standard \emph{SolarSoft} procedures. The
following procedures were applied in this order: decompression of
binary data, dead-time correction, flat-fielding, local-gain
correction, and correction for the geometrical distortion of the
detector. Finally, the data were calibrated to radiometric units
using the radiometry procedure. Technical details about the
instrument, corrections, and calibration procedures are available
in the SUMER Data Cookbook \citep{cit_sumdatacb} and references
therein. After the radiometric calibration, the obtained specific
intensities were multiplied by a factor of five to compensate for
the reduction of the observed signal due to the partially closed
aperture door. The precision of the SOHO/SUMER absolute
radiometric calibration was maintained to within 15\% during the
early years of operations
\citep[][]{1997SoPh..170...75W,1998ApOpt..37.2646S}; however, it
may have decreased since. Due to the narrow width (0.28~arcsec) of
the slit used for the observations, the influence of the
instrumental profile is small. We determined that the
deconvolution of the instrumental profile modifies the observed
Lyman-$\alpha$ intensities by up to 5\%. As this is below the
overall uncertainties of the radiometric calibration, we used here
the Lyman-$\alpha$ profiles as observed. Due to uncertainties in
the SOHO/SUMER pointing precision, the solar $X$ and $Y$
coordinates referred to throughout this work may be shifted by up
to $\pm15$~arcsec. However, such shifts do not considerably affect
our results.
\begin{figure*}
\centerline{\includegraphics[width=15cm]{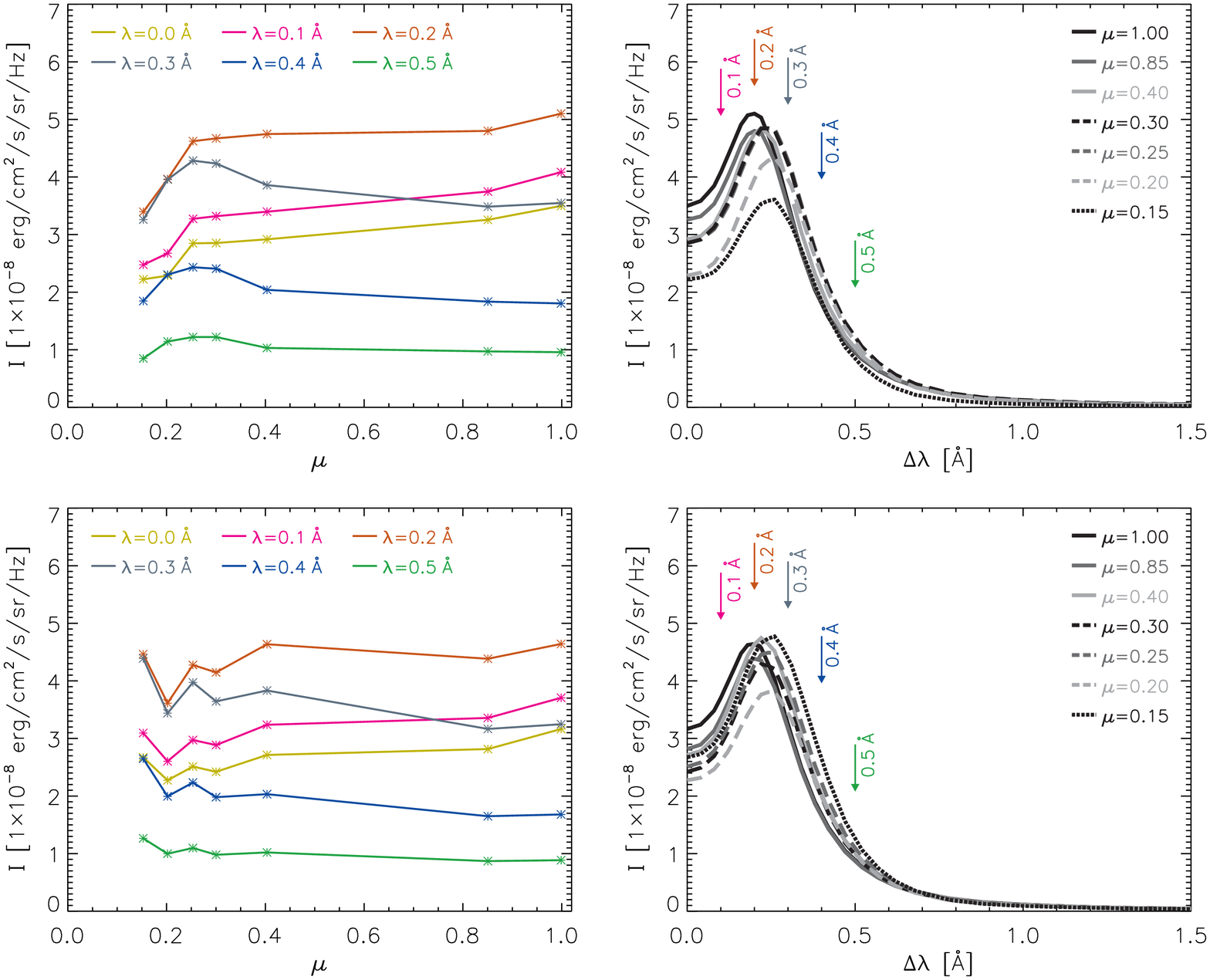}}
\caption{Left column: Spatially averaged Lyman-$\alpha$
intensities for selected wavelengths (marked in the upper left
corner) as a function of $\mu$. Right column: Spatially averaged
Lyman-$\alpha$ symmetrized profiles obtained from regions
corresponding to Fig.~\ref{fig:Rasters}. The top row shows the
data obtained on June 24, 2008 and the bottom row the data
obtained on June 25, 2008.} \label{fig:mu+I}
\end{figure*}
\begin{figure*}
\centerline{\includegraphics[width=14cm]{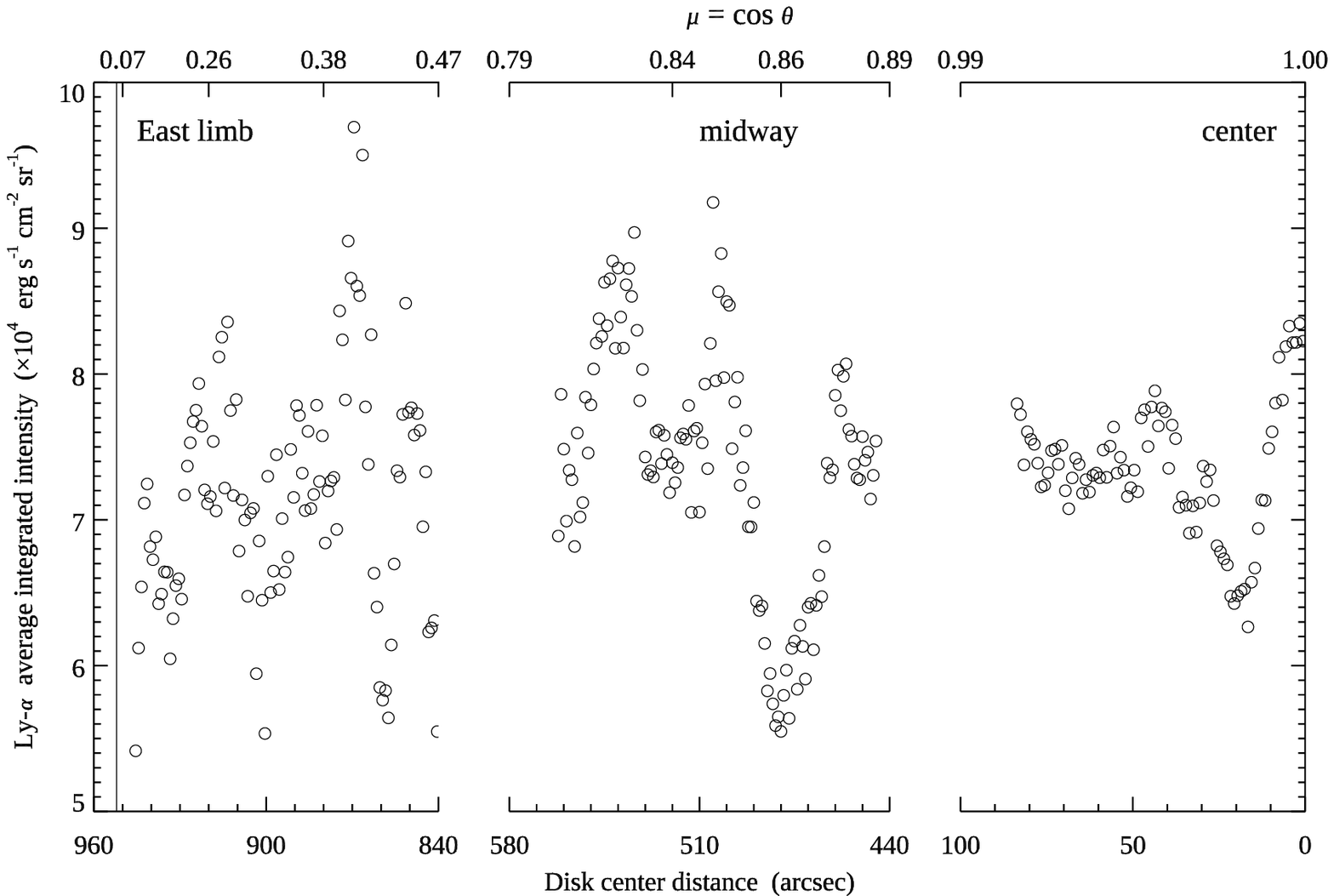}}
\caption{Centre-to-limb variation of the Lyman-$\alpha$
wavelength-integrated intensities averaged over concentric arcs
with a width of 1 arcsec. Each plotted circle in sections centre
and midway represents an average from rasters obtained on June 24,
25, and 26, 2008. For the near-limb section, only rasters from
June 24 and 25 were used. The thin vertical line shows the solar
radius of 952 arcsec as seen from SOHO.} \label{fig:arc_average}
\end{figure*}


\subsection{Analysis of the centre-to-limb variation}\label{Sect:averaged1}

\begin{figure*}
\centerline{\includegraphics[width=16.0cm]{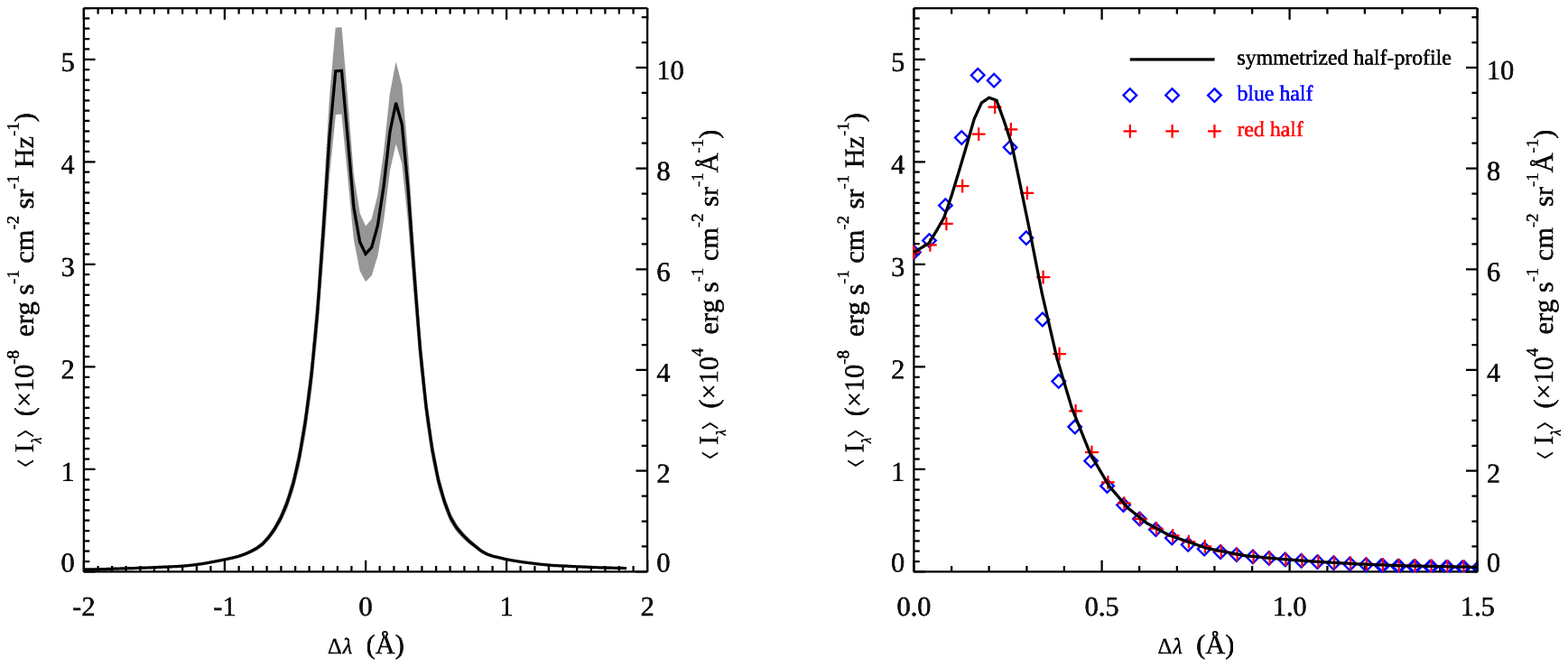}}
\caption{Left panel: Reference quiet-Sun
Lyman-$\alpha$ profile obtained as an average over eight observed
SOHO/SUMER rasters. The grey area indicates the uncertainty of the
reference profile estimated to be $\pm$(15/$\sqrt{3}$)\%. Right panel: Half of the symmetrized reference
Lyman-$\alpha$ profile with an indicated shape of the blue and the
red part of the full profile.} \label{fig:reference}
\end{figure*}
To analyse the variation of the Lyman-$\alpha$ profiles across the
solar disk, we selected seven regions in the rasters obtained on
June 24 and 25 because the quiet-Sun limb observations were only
performed on these days. In the selected regions (see
Fig.~\ref{fig:Rasters}), we averaged the specific Lyman-$\alpha$
intensity at each wavelength. We plot the averaged intensities for
selected wavelengths in the left-hand column of
Fig.~\ref{fig:mu+I}. The regions centred at $\mu$ equal to 1.0,
0.85, and 0.4 encompass large portions of the observed rasters.
The averaging over such wide areas minimizes the influence of the
significant local variations that are clearly demonstrated in all
rasters and are shown in Fig.~\ref{fig:Rasters} by the presence of
stochastically distributed bright and dark structures. Such a
large local variability of the Lyman-$\alpha$ spectra on the solar
disk is caused by the significant differences between the network
and inter-network regions. The effect of local intensity
variations is apparent in the averaged intensities from the
regions centred at $\mu$ equal to 0.3, 0.25, 0.2, and 0.15. These
regions cover considerably smaller portions of the observed
rasters and were selected to investigate the variation of the
Lyman-$\alpha$ intensity close to the solar limb. The average
intensities from these smaller regions vary considerably between
neighbouring regions (see the left-hand column of
Fig.~\ref{fig:mu+I}). In the right-hand column of
Fig.~\ref{fig:mu+I}, we plot the averaged Lyman-$\alpha$ line
profiles obtained from each region; we only plot half of the
symmetrized Lyman-$\alpha$ profile. In general, the Lyman-$\alpha$
profiles observed on the solar disk are typically asymmetric, with
the dominant blue peak (see e.g. \citet{2008A&A...492L...9C}).
However, the variation of the intensities in the blue and red
parts of the Lyman-$\alpha$ profile as a function of $\mu$ is very
similar, with only a small difference ($\sim$10\%) in some
wavelengths. From Fig.~\ref{fig:mu+I} (right-hand column), it is
clear that the profiles obtained at different regions vary.
However, the available set of observed data is not broad enough
for a statistically significant analysis of these
variations.\vspace{0.2 cm}

To assess in more detail the centre-to-limb variation of the
Lyman-$\alpha$ integrated intensity, we averaged the observed
data obtained on all three days over concentric arcs with a width
of 1 arcsec. The results (see Fig.~\ref{fig:arc_average}) clearly
show that the local variability of the integrated Lyman-$\alpha$
intensity is greater than any perceptible trend in the
centre-to-limb variation. This is in agreement with the findings
of \citet{2008A&A...492L...9C}.


\subsection{Full-disk Lyman-$\alpha$ profile}\label{Sect:averaged2}

Due to the absence of a clear centre-to-limb variation trend, we
derived the full-disk Lyman-$\alpha$ profile by spatially
averaging the specific intensities at each wavelength over all
eight observed rasters. In this way, we obtained the reference
quiet-Sun Lyman-$\alpha$ profile that represents the minimum of
solar activity. The integrated intensity of the reference
Lyman-$\alpha$ profile is 7.274 erg cm$^{-2}$ s$^{-1}$ str$^{-1}$.
We show the reference Lyman-$\alpha$ profile in the left-hand
panel of Fig.~\ref{fig:reference}, together with its
uncertainties. In the estimation of the uncertainties, we took
into account the findings of \citet{1997SoPh..170...75W}, who
attribute the $\pm15$\% uncertainty of SOHO/SUMER radiometric
calibration to the day-to-day variation of the flat field. Thus,
according to a general formula for error propagation, averaging
over three days should lead to a reduction of the uncertainties
from the admittedly optimistic 15\% to $\pm$(15/$\sqrt{3}$)\%. In
the right-hand panel of Fig.~\ref{fig:reference}, we plot one half
of the symmetrized reference Lyman-$\alpha$ profile. In
Appendix~\ref{App:Lya}, we provide the data of both the full and
symmetrized reference Lyman-$\alpha$ profiles. The same data sets
are attached to this paper in an online accessible form.


\section{Variation with the solar cycle}\label{Sect:cycle}

\subsection{Lyman-$\alpha$ line}\label{Sect:cycle_Lya}

\begin{figure*}
\centerline{\includegraphics[width=13.5cm]{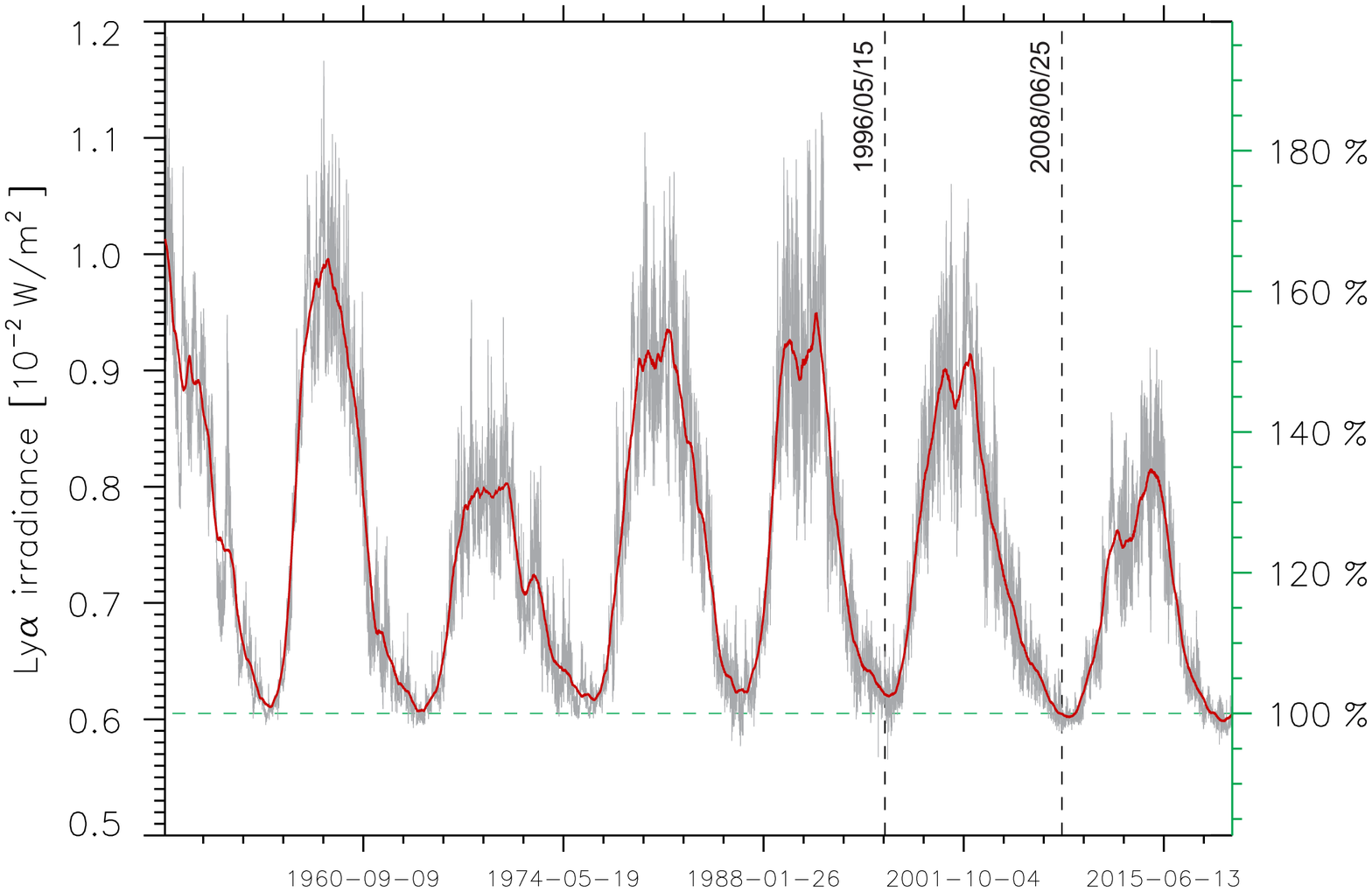}}
\caption{Variation of the total Lyman-$\alpha$ irradiance
based on the LISIRD composite Lyman-$\alpha$ index (grey line).
The red line shows the same index smoothed by a running average over
400 days. Dashed vertical lines mark the reference date for the
observations used in the current work (June 25, 2008) and the date
(May 15, 1996) of the observations used in
\citet{1998ApJS..119..105W}. The $y$-axis on the right-hand side
gives the values of the index as a percentage of the value on the
reference day ($0.605\times10^{-2}$ W m$^{-2}$), marked by the
horizontal dashed line.} \label{fig:variation}
\end{figure*}

The Lyman-$\alpha$ radiation from the solar disk is not constant
over time but varies significantly with the solar cycle. To take
these changes into account, we used the LISIRD composite
Lyman-$\alpha$ index \citep[][]{2019E&SS....6.2263M}. The value of
this index, shown in Fig.~\ref{fig:variation}, clearly follows the
solar cycle pattern and can vary considerably -- extreme differences
between the minima and maxima of a cycle can reach up to 100\%. The
same index was also used by \citet{2015A&A...581A..26L} to
investigate the variation of the Lyman-$\alpha$ spectra throughout
solar cycle 23. \vspace{0.2 cm}

In the present work, we used the LISIRD composite Lyman-$\alpha$
index smoothed by the running mean over 400 days (13 months),
which is usually used for the representation of solar cycle minima
and maxima \citep[see e.g.][]{2015LRSP...12....4H}. The typical
differences between the minima and maxima of the smoothed index
(red line in Fig.~\ref{fig:variation}) are around 50\%. We used
the smoothed index to derive coefficients describing the change in
the Lyman-$\alpha$ illumination between the date when the
reference Lyman-$\alpha$ profile was obtained (June 25, 2008) and
the date when the observations analysed by the radiative transfer
modelling were obtained. These coefficients can be derived for any
date as a ratio between the value of the Lyman-$\alpha$ index on
the day of observation and the value of the index on the reference
day (June 25, 2008), which is $0.605\times10^{-2}$ W m$^{-2}$. In
Appendix~\ref{App:cycle}, we list coefficients (obtained for the
smoothed index) for selected days from the start of the SOHO
observations until 2020. These coefficients are also provided
 as an online attachment. We note that we applied the same
coefficients to the specific intensities throughout the
Lyman-$\alpha$ profile. This is in agreement with the findings of
\citet{2018GeoRL..45.2138K}, whose empirical model shows a
negligible wavelength dependence of the variation of
Lyman-$\alpha$ irradiance with the solar cycle. We note, however,
that the phenomenological model of the solar Lyman-$\alpha$ line
from \citet{2018ApJ...852..115K} points to a more pronounced
wavelength dependence of the Lyman-$\alpha$ irradiance on the
evolution of the solar activity.

\subsection{Higher Lyman lines}\label{Sect:cycle_Ly_other}

While the Lyman-$\alpha$ line is the most intense source of
illumination for the hydrogen radiative transfer modelling of
chromospheric and coronal structures, the higher Lyman lines also
play an important role. In the absence of a better source of
information about their cyclic variations, we again used the
LISIRD composite Lyman-$\alpha$ index as a proxy for the variation
of higher Lyman lines. We note that the findings of
\citet{2012A&A...542L..25L} show that the ratio of the irradiance
in Lyman-$\alpha$ and Lyman-$\beta$ lines is not constant over
time. However, the differences in the ratio of these two lines
appear to be considerably smaller than the variation of the
irradiance over the solar cycle (see \citet{2012A&A...542L..25L}
for more details). We thus assume here a constant ratio between
Lyman-$\alpha$ and higher Lyman lines.\vspace{0.2 cm}

In the present work, we did not derive the solar disk spectra in
the Lyman lines above Lyman-$\alpha$. We used the reference spectra
provided by \citet{1998ApJS..119..105W}. The data used by
\citet{1998ApJS..119..105W} were also obtained by SOHO/SUMER on
May 15, 1996, also within a period of minimum activity of solar
cycle 22. However, there is a small difference (see
Fig.~\ref{fig:variation}) in the irradiance between the date used
by \citet{1998ApJS..119..105W} and the date of the Lyman-$\alpha$
observations used in the present work (June 25, 2008). Therefore,
we first used a coefficient (0.973) to normalize the intensity on
these two dates, and then we applied the same coefficients as for
the observations analysed in the present work. In
Appendix~\ref{App:cycle}, we list the combined coefficients for
the transformation of \citet{1998ApJS..119..105W} reference
profiles to the selected dates.

\subsection{H$\alpha$ line}\label{Sect:cycle_Ha}

Another important source of illumination of chromospheric and
coronal structures is the Balmer H$\alpha$ line. However, the
intensity of this line, in contrast to the lines of the Lyman
series, does not change significantly with the solar cycle. For
example, the MuSIL Corrected SORCE SIM
index\footnote{\url{lasp.colorado.edu/lisird/data/musil_sim}} in
the wavelength range centred at 6562\,$\AA$ shows very small
differences (less than 1\%) between the minima and maxima of the
H$\alpha$ intensity. We can, therefore, assume the same H$\alpha$
incident-radiation profile throughout the solar cycle. The
reference profiles from \citet{1961ZA.....53...37D} together with
the disk continuum intensities from, for example,
\citet{1976asqu.book.....A} are typically used for all Balmer
lines.


\section{Influence of the incident-radiation change on synthetic profiles}\label{Sect:influence}

\renewcommand\arraystretch{1.3}
\begin{table*}
  \caption{List of specific intensities in
  the peak of symmetrized Lyman-$\alpha$, Lyman-$\beta$, and Lyman-$\gamma$ incident-radiation profiles adapted to the selected dates that follow the
change in illumination with the solar cycle. In the last column,
we show relative differences between the intensities on the given
date and the reference date June 25, 2008. }
  \normalsize
  \centerline{
  \begin{tabular}{ccccc}
      \toprule
      \multirow{2}{*}{Date} & Lyman-$\alpha$ peak intensity & Lyman-$\beta$ peak intensity & Lyman-$\gamma$ peak intensity & Difference \\
        & erg cm$^{-2}$ s$^{-1}$ str$^{-1}$ Hz$^{-1}$ & erg cm$^{-2}$ s$^{-1}$ str$^{-1}$ Hz$^{-1}$ & erg cm$^{-2}$ s$^{-1}$ str$^{-1}$ Hz$^{-1}$ & from $2008/06/25$ \\
      \midrule
      $2008/06/25$ & $4.74\times10^{-8}$ & $5.15\times10^{-10}$ & $1.23\times10^{-10}$ & 0 \% \\
      $2010/06/24$ & $5.06\times10^{-8}$ & $5.51\times10^{-10}$ & $1.31\times10^{-10}$ & 7 \% \\
      $2011/01/01$ & $5.28\times10^{-8}$ & $5.75\times10^{-10}$ & $1.37\times10^{-10}$ & 12 \% \\
      $2011/06/24$ & $5.67\times10^{-8}$ & $6.16\times10^{-10}$ & $1.47\times10^{-10}$ & 20 \% \\
      $2013/06/24$ & $5.99\times10^{-8}$ & $6.51\times10^{-10}$ & $1.55\times10^{-10}$ & 26 \% \\
      $2014/06/24$ & $6.37\times10^{-8}$ & $6.93\times10^{-10}$ & $1.65\times10^{-10}$ & 34 \% \\
      \toprule
  \end{tabular}}
  \label{tab:IncRad}
\end{table*}
\begin{table*}
  \caption{List of relative differences between
the central, integrated, and peak intensities of synthetic
profiles obtained by {\sc Model1} for the selected dates and the
intensities obtained for the reference date (June 25, 2008). For
each spectral line and analysed profile characteristic, we list
the maximum, median, and minimum differences.}
  \normalsize
  \centerline{
  \begin{tabular}{ccccccccccccc}
      \toprule
      \multirow{2}{*}{Date} & \multicolumn{3}{c}{Lyman-$\alpha$ intensity} & \multicolumn{3}{c}{Lyman-$\beta$ intensity} & \multicolumn{3}{c}{Lyman-$\gamma$ intensity} & \multicolumn{2}{c}{H$\alpha$ intensity} &  \\
        & centre & integral & peak & centre & integral & peak & centre & integral & peak & centre & integral \\
      \midrule
        & 7 \% & 6 \% & 5 \% & 6 \% & 6 \% & 3 \% & 6 \% & 6 \% & 2 \% & 3 \% & 3 \% & max\\
      $2010/06/24$  & 6 \% & 5 \% & 2 \% & 5 \% & 3 \% & 1 \% & 3 \% & 2 \% & 1 \% & 1 \% & 1 \% & median\\
        & 6 \% & 2 \% & - & 2 \% & 1 \% & - & 1 \% & 1 \% & - & - & - & min\\
        &&\\
       & 11 \% & 11 \% & 7 \% & 10 \% & 10 \% & 5 \% & 10 \% & 9 \% & 4 \% & 5 \% & 5 \% & max\\
       $2011/01/01$ & 10 \% & 9 \% & 3 \% & 8 \% & 5 \% & 2 \% & 5 \% & 3 \% & 2 \% & 2 \% & 2 \% & median\\
        & 10 \% & 3 \% & 1 \% & 3 \% & 2 \% & 1 \% & 1 \% & 1 \% & 1 \% & - & - & min\\
        &&\\
       & 19 \% & 18 \% & 13 \% & 18 \% & 17 \% & 9 \% & 16 \% & 16 \% & 6 \% & 8 \% & 9 \% & max\\
      $2011/06/24$  & 17 \% & 15 \% & 5 \% & 14 \% & 8 \% & 4 \% & 9 \% & 5 \% & 3 \% & 4 \% & 3 \% & median\\
        & 16 \% & 5 \% & 1 \% & 6 \% & 4 \% & 1 \% & 2 \% & 2 \% & 2 \% & - & - & min\\
        &&\\
       & 25 \% & 24 \% & 17 \% & 24 \% & 22 \% & 13 \% & 22 \% & 21 \% & 9 \% & 11 \% & 11 \% & max\\
       $2013/06/24$ & 23 \% & 20 \% & 7 \% & 19 \% & 11 \% & 5 \% & 12 \% & 7 \% & 4 \% & 5 \% & 5 \% & median\\
        & 22 \% & 6 \% & 2 \% & 8 \% & 5 \% & 2 \% & 3 \% & 3 \% & 3 \% & 1 \% & - & min\\
        &&\\
       & 33 \% & 32 \% & 22 \% & 31 \% & 29 \% & 16 \% & 29 \% & 28 \% & 11 \% & 14 \% & 15 \% & max\\
       $2014/06/24$ & 30 \% & 26 \% & 10 \% & 25 \% & 15 \% & 7 \% & 16 \% & 9 \% & 6 \% & 6 \% & 6 \% & median\\
        & 29 \% & 8 \% & 3 \% & 11 \% & 7 \% & 2 \% & 4 \% & 3 \% & 3 \% & 1 \% & - & min\\
      \toprule
  \end{tabular}}
  \label{tab:influence}
\end{table*}
\renewcommand\arraystretch{1.0}
\begin{figure*}
\centerline{\includegraphics[width=13cm]{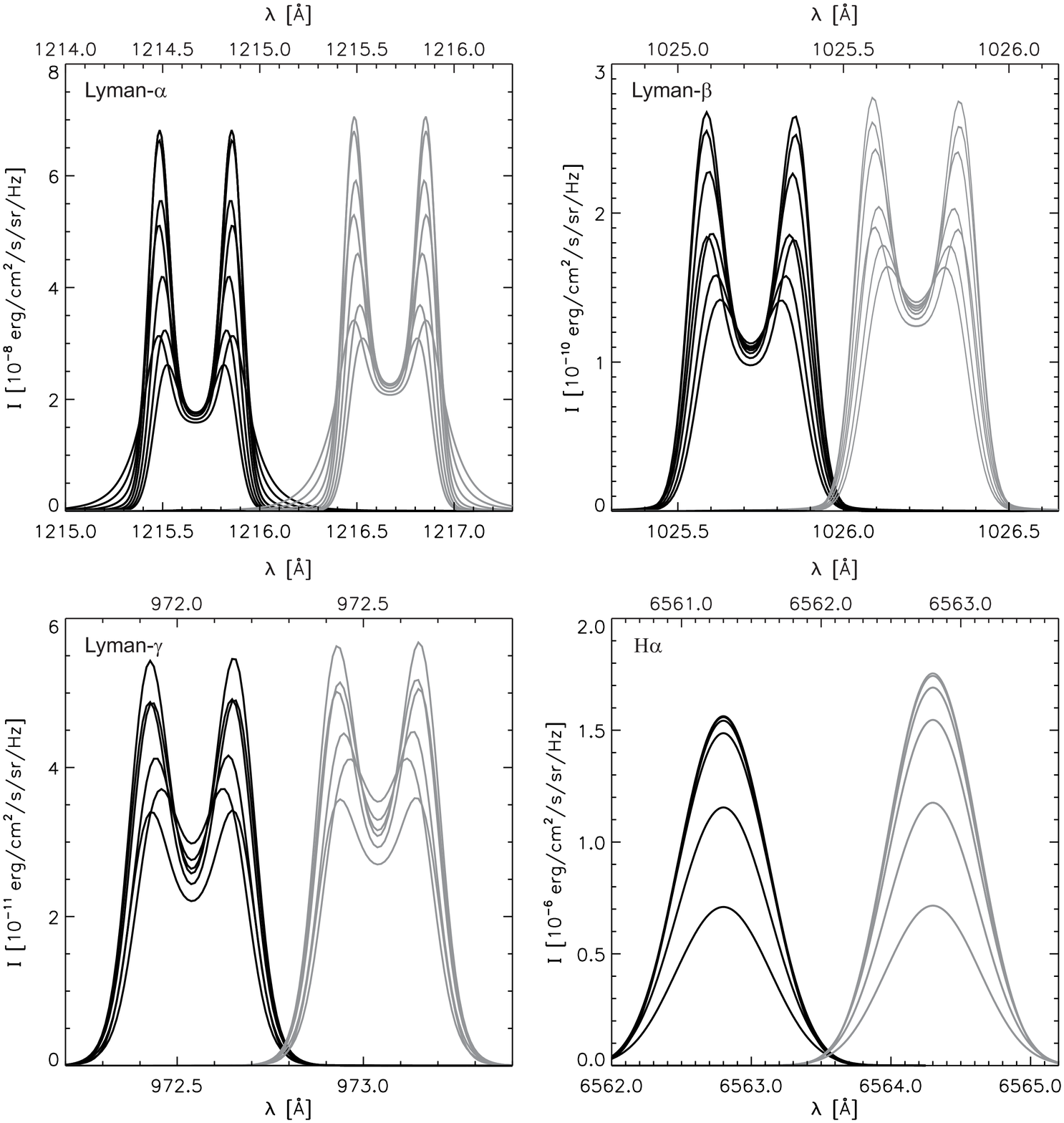}}
\caption{Comparison of synthetic spectra produced by {\sc
Model1} in the single-thread configuration with the illumination
data for the reference date June 25, 2008 (black lines) and the
date with the maximum illumination change June 24, 2014 (grey
lines). We plot only the reversed profiles obtained at different
positions along the length of the used 2D prominence fine
structure model. We note that the bottom $x$-axis in each panel
corresponds to the black profiles and the top $x$-axis to the grey
profiles. All plotted profiles are convolved with the instrumental
profiles of SOHO/SUMER (Lyman lines) or Meudon/MSDP (H$\alpha$).}
\label{fig:minVSmax}
\end{figure*}
In the previous section, we discussed the fact that the solar
radiation in Lyman lines is not constant over time but changes
significantly with the solar cycle. We have also presented a
method that can be used to adapt the reference incident radiation
profiles to a specific date when a set of observations analysed by
radiative transfer modelling was obtained. Such a method can lead
to an increased accuracy of the boundary conditions of radiative
transfer models. \vspace{0.2 cm}

To estimate the extent of the influence of the illumination changes on the results of radiative transfer models, we used the 2D non-LTE model of vertical prominence fine structures
developed by \citet{2001A&A...375.1082H}. This model was used for
the analysis of prominence spectral observations by, for example,
\citet{2010A&A...514A..43G,2012A&A...543A..93G,2014A&A...567A.123G}.
In the present work, we adopted a set of model input parameters
derived as a good fit to observations of a quiescent prominence by
\citet{2010A&A...514A..43G}, referred to as {\sc Model1} therein
and hereafter. This allowed us to analyse the influence of the
modified incident radiation intensities on a model with a
realistic set of input parameters \citep[for more details on the
parameters of the model, see Table 2 of][]{2010A&A...514A..43G}. We
note that the analysis presented here serves only as an estimate
of the influence of varying illumination on the results of
radiative transfer models. A more in-depth investigation involving
models of different solar structures would be needed to reveal the
true extent of this influence. However, such an investigation is
beyond the scope of the current paper.\vspace{0.2 cm}

To follow the cyclic change in the solar disk radiation in Lyman
lines, we selected five ad hoc dates representing the
increase in the Lyman-$\alpha$ intensity from the minimum of the
cycle to its maximum. The specific intensities in the peak of
symmetrized Lyman-$\alpha$, Lyman-$\beta$, and Lyman-$\gamma$
incident radiation profiles for these dates are listed in
Table~\ref{tab:IncRad}. The reference date is the day of
observation used in the present work (June 25, 2008), which
represents a minimum of the solar activity. For each date, we
computed {\sc Model1} in the single-thread configuration and
produced the synthetic spectra of the Lyman lines and the
H$\alpha$ line. The 2D prominence fine structure model used here
produces synthetic line profiles with a variety of shapes and
intensities, depending on the orientation of the line-of-sight and
its position along the length (or width) of the model. More
details are available in, for example, Sect.~5 of \citet{2005A&A...442..331H}
or Sect.~4 of \citet{2007A&A...472..929G}. In the present work, we
took into account the synthetic profiles obtained with the
line-of-sight perpendicular to the longer dimension of the model,
that is to say, perpendicular to the magnetic field. Only the profiles with
intensities higher than 20\% of the maximum intensity in each
spectral line are used. For consistency with observations, the
resulting synthetic spectra were convolved with the instrumental
profile of SOHO/SUMER \citep[see e.g. Sect.~3.1
of][]{2008A&A...490..307G} and of the Meudon/MSDP (Multichannel
Subtractive Double Pass) spectrograph working at the Meudon Solar
Tower \citep[][]{1991A&A...248..669M}.\vspace{0.2 cm}

We used three characteristics of the synthetic profiles for the
analysis of the effect of the illumination change. These are the
specific intensity in the centre of the line, its integrated
intensity, and the maximum specific intensity in the peaks of the
reversed profiles. We took into account only reversed profiles
with peak intensities of more than 120\% of the corresponding
line-centre intensity. We plot these reversed profiles in
Fig.~\ref{fig:minVSmax} to illustrate the changes in the synthetic
profiles due to modified incident radiation intensities. In this
figure, we compare the synthetic profiles for the reference date
June 25, 2008 (black lines) and the profiles for June 24, 2014
(grey lines). In Table~\ref{tab:influence}, we list the relative
differences between the central, integrated, and peak intensities
obtained for the selected dates and those obtained for the
reference date. For each date, spectral line, and analysed profile
characteristic, we list the maximum, median, and minimum
differences. We note that we do not list the peak intensities for
the H$\alpha$ line because the obtained synthetic profiles are not
reversed.\vspace{0.2 cm}

Table~\ref{tab:influence} clearly shows that the centre of the
Lyman-$\alpha$ line is strongly affected by the change in the
incident radiation. In fact, the relative differences between the
synthetic spectra obtained with the incident radiation data for
June 24, 2014, and those obtained for the reference date are
practically the same, as is the amplitude of change of the
incident radiation itself. Moreover, these differences vary only
slightly between the maximum and minimum values (see the second
column of Table~\ref{tab:influence}). The maximum differences in
the centre of Lyman-$\beta$ and Lyman-$\gamma$ lines are similar
to the differences in the centre of Lyman-$\alpha$. However, the
variation between the maximum and minimum differences is more
pronounced in these lines. The maximum relative differences
between the integrated intensities of all three Lyman lines are
also comparable to the amplitude of change in the incident
radiation. The modification of the incident radiation has the
least impact on the specific intensity in the peaks of the
reversed profiles of Lyman lines. However, even here the peak
intensities may change by up to two-thirds of the amplitude of
change of the incident radiation. Only the most intense peaks are
practically unaffected by the change in  the incident radiation.
This contrast between the influence of the modified illumination
on the centre and the peaks of synthetic profiles can be seen in
Fig.~\ref{fig:minVSmax}. Interestingly, the central and integrated
intensities of the H$\alpha$ line can also be significantly
affected by the change in illumination in the Lyman lines. This is
despite the fact that the illumination in the H$\alpha$ line
itself is constant (see Sect.~\ref{Sect:cycle_Ha}). Both the
specific intensity in the H$\alpha$ line centre and its integrated
intensity are typically affected by around one-fifth of the
amplitude of change of the incident radiation in Lyman lines.
However, this influence can increase to up to two-fifths in the
most intense H$\alpha$ profiles. \vspace{0.2 cm}

We also performed a further assessment of the influence of
changing illumination on the modelling results, this time using
the multi-thread configurations of {\sc Model1} with randomly
assigned line-of-sight velocities \citep[see][for more
details]{2010A&A...514A..43G}. When we used a multi-thread model
with ten identical threads (each with the input parameters of {\sc
Model1}) and the line-of-sight velocities selected randomly from an interval
$\pm$10 km/s, we found that, depending on the actual setup of
line-of-sight velocities and the positional shifts in the
multi-thread model, the synthetic profiles were affected to a
slightly lesser degree than those produced by the single-thread
model.


\section{Discussion}\label{Sect:discussion}

The integrated intensity of the reference quiet-Sun Lyman-$\alpha$
profile from Fig.~\ref{fig:reference} ($7.36\times10^{4}$ erg
cm$^{-2}$ s$^{-1}$ str$^{-1}$) translates to a total
Lyman-$\alpha$ irradiance of $0.501\times10^{-2}$ W m$^{-2}$.
However, this value is about 20\% lower than the value of the
LISIRD composite Lyman-$\alpha$ index on the reference date (June
25, 2008), which is $0.605\times10^{-2}$ W m$^{-2}$. This
discrepancy can be caused by a combination of various factors. The
first is the uncertainty in the calibration of the SOHO/SUMER data
used here, which is $\pm$(15/$\sqrt{3}$)\%, together with the
uncertainty of the LISIRD composite Lyman-$\alpha$ index
calibration. The estimated uncertainty of the value of the index
on the reference date is around 6\% \citep[see][for more
details]{2019E&SS....6.2263M}. The second factor is the presence
of an active region and coronal holes at each pole during the
dates when SOHO/SUMER observations were obtained (see
Fig.~\ref{fig:context}). Both these features affect (and probably
increase) the total Lyman-$\alpha$ irradiance represented by the
LISIRD composite Lyman-$\alpha$ index, but they do not affect the
averaged quiet-Sun reference profile derived from the SOHO/SUMER
raster scans. To quantify the extent of the influence of active
regions, plages, and coronal holes on the Lyman-$\alpha$
irradiance, we will need to analyse detailed Lyman-$\alpha$
spectral observations of these features. However, such an analysis
is beyond the scope of the current paper. \vspace{0.2 cm}

For the assessment of the influence of the varying solar
illumination on the synthetic spectra, we used the LISIRD
composite Lyman-$\alpha$ index smoothed by a running averaging
over 400 days (see Sect.~\ref{Sect:cycle_Lya}). However, from
Fig.~\ref{fig:variation}, it is clear that the variations in the
smoothed index are considerably smaller than the variations in the
actual, daily-averaged values of the index itself. As we
demonstrated in the previous section, the change in incident
radiation can have a significant effect on the synthetic spectra.
Therefore, for applications where it is necessary to employ the
most accurate incident radiation data, we would suggest using the
actual values of the composite Lyman-$\alpha$ index or other
shorter-term averaging. We note, however, that the use of a
running mean over a period other than 400 days will result in
coefficients that vary from those listed in
Appendix~\ref{App:cycle}. In the case when the actual,
daily-averaged values of the Lyman-$\alpha$ index are used, it is
important to take into account the fact that these values
correspond to the radiation from the solar disk as seen from
Earth. Therefore, the actual values should be used for the
modelling of on-disk chromospheric or coronal structures that are
not too distant from the disk centre. On the other hand, the limb
structures, such as prominences, spicules, or CME cores, are
illuminated only by a portion of the solar disk visible from the
Earth and by a portion of the disk beyond the limb. However, the
radiation from beyond the limb is obviously not included in the
actual values of the LISIRD composite Lyman-$\alpha$ index.
Therefore, special care should be exercised when these values are
used for the analysis of the limb observations. \vspace{0.2 cm}

Realistic determination of the Lyman-$\alpha$ incident radiation
profile is critically important for the diagnostics of the physical
properties of the chromospheric and coronal structures with
prominence-like conditions. This is because the Lyman-$\alpha$
line profiles emerging from these structures are produced by
the combined effect of their physical conditions, the illumination
from the solar surface, and the properties of the spectral-line
formation mechanism in the regime of partial redistribution (PRD).
For isothermal-isobaric prominence slab models,
\citet{1987A&A...183..351H} demonstrated that the emergent
Lyman-$\alpha$ profile substantially differs when modelled with
realistic PRD or with approximate complete redistribution (CRD).
This is because, at low prominence densities, the photon
scattering plays a dominant role in the line formation. This
scattering is relatively well described by CRD only in the
Lyman-$\alpha$ line core up to approximately three Doppler widths,
while PRD is needed in the wings. Under typical prominence
temperatures, three Doppler widths amount to about 0.15 \AA,\ while
the peaks of the Lyman-$\alpha$ incident radiation profile are located
around 0.2 \AA\ from the line centre (see
Fig.~\ref{fig:reference}). This means that the peaks of the
incident radiation profile are partially reproduced by
quasi-coherent scattering (which is not the case of CRD). As a
result, the line-core intensity is almost the same for both PRD
and CRD, but the peaks are produced predominantly in the PRD
regime. Moreover, far wings are much lower in the PRD case
compared to CRD, which is a well-known effect in stellar
atmospheres. Furthermore, it is interesting to note that the
Lyman-$\alpha$ line-centre intensity in the 1D models is
practically equal to the incident line-centre intensity multiplied
by the geometrical dilution factor
\citep[see][]{1987A&A...183..351H}. This is also true in the 2D
prominence fine structure model used here. The aspects of the
Lyman-$\alpha$ line formation mechanism described above have
several consequences related to the shape and intensity of the
incident radiation profile. First, the peaks of the synthetic
Lyman-$\alpha$ profiles depend on the peaks of the incident
radiation profile because they are quasi-coherently reproduced.
However, the degree of the reproduction depends mainly on the
electron density, which is a key parameter for the determination of
departures from CRD. At high densities, the line formation
mechanism in the peaks would approach the CRD regime. However,
under the prominence-like conditions, the electron density is
relatively low and the formation mechanism is in the PRD regime.
Second, the synthetic intensity in the Lyman-$\alpha$ line core
depends on the temperature and density structure of the
prominence-corona transition region (PCTR), where, in the case of
prominence-like structures, Lyman-$\alpha$ is formed due to its
very high optical thickness. The dependence is caused by the
thermal component of the Lyman-$\alpha$ line source function, which
becomes important at higher temperatures. Therefore, any
unrealistic determination of the line-core intensity of the
Lyman-$\alpha$ incident radiation profile significantly decreases
the accuracy of the plasma temperature diagnostics. We note that
similar effects also take place in the case of the Lyman-$\beta$ line,
but the formation of this line is less sensitive to PRD due to the
strong coupling with the H$\alpha$ line, which is formed in the CRD
regime. \vspace{0.2 cm}

The Lyman-$\alpha$ profile observed on the solar disk is in
general asymmetric, with a stronger blue peak, as was shown by
\citet{2008A&A...492L...9C} and
\citet{2009A&A...504..239T,2009ApJ...703L.152T}. These authors
also showed that the Lyman-$\beta$ line profile obtained on the
solar disk generally shows an opposite asymmetry, with a stronger
red peak. This behaviour might be caused by differential flows in
the solar atmosphere due to a mechanism similar to that
demonstrated by \citet{2008A&A...490..307G} in prominences.
However, detailed causes of these opposing asymmetries of
Lyman-$\alpha$ and higher Lyman lines are not yet clearly
understood. The reference full-disk Lyman-$\alpha$ profile
presented here exhibits the same stronger-blue-peak asymmetry.
Such an asymmetry might play a negligible role in a majority of
radiative transfer models. Indeed, many of the current models
assume symmetric incident radiation profiles -- the reason for which
we also provide here the symmetrized reference Lyman-$\alpha$
profile (Fig.~\ref{fig:reference}). However, the true influence of
the asymmetries of the incident radiation profiles will need to be
investigated in the future in greater detail. On the other hand, it is
important to note that asymmetries in the incident radiation
profiles may play a more significant role in the radiative
transfer of polarized light, especially when the 3D polarized
light radiative transfer is considered \citep[see e.g. the 3D
polarized radiative transfer code PORTA
by][]{2013A&A...557A.143S}. This is because the polarization is
sensitive to the breaking of symmetries. However, the results of
polarized light radiative transfer modelling might be even more
sensitive to the local inhomogeneities of the illumination from
the solar disk than to the asymmetries of the incident radiation
profiles.


\section{Conclusions}\label{Sect:conclusions}

We derived the reference quiet-Sun Lyman-$\alpha$ spectral
profile, which represents the Lyman-$\alpha$ radiation from the
solar disk during the minimum of the solar activity. This profile
(Fig.~\ref{fig:reference}) can serve as a representative
illumination (incident radiation) boundary condition for radiative
transfer models of chromospheric and coronal structures, such as
prominences, spicules, chromospheric fibrils, solar wind, and CME
cores. Moreover, solar Lyman-$\alpha$ profile and the total
Lyman-$\alpha$ irradiance are important for the Earth's
ionosphere, the heliosphere, and the atmospheres of planets,
moons, and comets.\vspace{0.2 cm}

To derive the reference Lyman-$\alpha$ profile, we used what is, to
our knowledge, the most comprehensive data set of SOHO/SUMER
Lyman-$\alpha$ observations of the solar disk obtained without the
use of the attenuator. This data set comprises eight
Lyman-$\alpha$ raster scans obtained in various quiet-Sun regions
over three consecutive days (June 24 - 26, 2008) with a reduced
incoming photon flux due to a partially closed aperture door of
the telescope. The rasters obtained on June 24 and 25 were
previously presented and analysed by \citet{2008A&A...492L...9C}.
We provide the reference Lyman-$\alpha$ profile in a tabular form
in Appendix~\ref{App:Lya} and online. As we discuss in
Sect.~\ref{Sect:averaged1}, the uncertainties in the determination
of the reference profile are $\pm$(15/$\sqrt{3}$)\%. These are
caused by the uncertainty of $\pm15$\% (at best) in the SOHO/SUMER
radiometric calibration
\citep[][]{1997SoPh..170...75W,1998ApOpt..37.2646S} alleviated by
the use of observations obtained on three days. \vspace{0.2 cm}

Solar radiation in Lyman lines is not constant over time but
varies significantly (up to 100\%) with the solar cycle (see
Sect.~\ref{Sect:cycle}). Such a dramatic change in the boundary
conditions of radiative transfer models can have a profound effect
on their results, as we show in this work. To take the variation
of Lyman-$\alpha$ with the solar cycle into account, we present in
Sect.~\ref{Sect:cycle_Lya} a method for adapting the incident
radiation Lyman line profiles to a specific date. This method uses
the LISIRD composite Lyman-$\alpha$ index
\citep[][]{2019E&SS....6.2263M} to derive coefficients for
the modification of the reference Lyman-$\alpha$ profile presented
here and the reference spectra of higher Lyman lines from
\citet{1998ApJS..119..105W}. The list of coefficients provided in
Appendix~\ref{App:cycle} (and online) covers the lifetime of
SOHO.\vspace{0.2 cm}

In the present work, we used the 2D non-LTE radiative transfer
model of prominence vertical fine structures
\citep{2001A&A...375.1082H} to analyse the influence of the change
in illumination in the Lyman lines on the synthetic spectra. We
used this model as an example that allowed us to estimate the impact
of the illumination change on synthetic spectra, which are in broad
agreement with prominence observations \citep[see][for more
details]{2010A&A...514A..43G}. However, a more thorough
investigation employing models of various chromospheric and
coronal structures will be needed to establish the true extent of
the influence of varying incident radiation boundary conditions on
the modelling results. Such an in-depth investigation is beyond
the scope of this paper. Nevertheless, our analysis clearly shows
that the modification of the incident radiation has a very
significant impact on the resulting synthetic spectra. This impact
is most prominent in the centre of the Lyman lines -- especially the
Lyman-$\alpha$ line -- and in their integrated intensities. There, a
change in the incident radiation can often translate into a large
change in these profile characteristics. The impact on the peak
intensities of the reversed profiles of Lyman lines is also
significant, albeit smaller. Only in the case of the most intense
profiles is the impact of the modified illumination on the
intensities in the peaks minimal. The change in illumination in
the Lyman lines also affects the H$\alpha$ line profiles to a
significant degree despite the fact that the illumination in the H$\alpha$
line does not change. The H$\alpha$ is affected because it is
formed by the transition from the third to the second level of
hydrogen. The populations of these levels are strongly affected by
radiation in the Lyman-$\alpha$ and Lyman-$\beta$
lines.\vspace{0.2 cm}

The strong sensitivity of the synthetic spectra to the incident
radiation shows that better observations and more detailed studies
of the solar disk radiation in the Lyman lines will be needed.
This strong sensitivity also means that a detailed forward-modelling
analysis of observations of chromospheric and coronal structures
will benefit from up-to-date incident radiation boundary
conditions. The incident radiation profiles adapted to the date of
the analysed observations should be especially considered in a
case when the analyzed observations are obtained during maxima of
solar activity. This is because the reference incident radiation
profiles, both those of \citet{1998ApJS..119..105W} and those
provided here, correspond to solar minima. However, careful
consideration should be given to the choice of the actual values
of the LISIRD composite Lyman-$\alpha$ index or other measures of
the variability of solar radiation in Lyman lines.


\begin{acknowledgements}

S.G. and P.H. acknowledge support from the grant No. 19-16890S and
grant No. 19-17102S of the Czech Science Foundation (GA \v CR).
S.G. acknowledges support from the grant No. 19-20632S of the
Czech Science Foundation (GA \v CR). S.G., P.S., P.H. and J.K.
acknowledge support from the Joint Mobility Project SAV-18-03 of
Academy of Sciences of the Czech Republic and Slovak Academy of
Sciences.  P.S. and J.K. acknowledge support from the project VEGA
2/0048/20 of the Science Agency. S.G. and P.H. thank for the
support from project RVO:67985815 of the Astronomical Institute of
the Czech Academy of Sciences. P.S. thanks for the support from
the grant No. 19-16890S of the Czech Science Foundation (GA \v
CR). P.S. and J.K. thank for the support from the grant No.
19-17102S of the Czech Science Foundation (GA \v CR). The SUMER
project is financially supported by DLR, CNES, NASA, and the ESA
PRODEX Programme. SUMER is part of SOHO of ESA and NASA. We
acknowledge the use of data obtained from the LASP Interactive
Solar Irradiance Data Center (LISIRD) available at
\url{lasp.colorado.edu/lisird}. The context H$\alpha$ images used
in this paper were obtained by the Big Bear Solar Observatory,
which is operated by the New Jersey Institute of Technology, and
by the SMART telescope at the Hida Observatory operated by Kyoto
University. The authors thank W. Curdt, J.-C. Vial, J. \v St\v
ep\'an and M. Exnerov\'a for valuable discussions.
\end{acknowledgements}



\begin{appendix}

\onecolumn
\section{The reference full-disk Lyman-$\alpha$ line profile}\label{App:Lya}

\renewcommand\arraystretch{1.3}
\begin{longtable}{ccc}
    \caption{Reference quiet-Sun Lyman-$\alpha$ profile (Fig.~\ref{fig:reference}, left) obtained as an average over eight
observed SOHO/SUMER raster scans obtained between June 24, 2008, and June 26, 2008. The central wavelength $\lambda_{0}$
    of the Lyman-$\alpha$ line is 1215.67\,$\AA$. The uncertainty of the
reference profile was estimated to be $\pm$(15/$\sqrt{3}$)\%.}\label{tab:App_fullProf}\\
    \toprule
    Wavelength ($\Delta\lambda_{0}$) [$\AA$] & Intensity [$10^{-8}\,\rm{erg}\,\rm{cm}^{-2}\,\rm{s}^{-1}\,\rm{sr}^{-1}\,\rm{Hz}^{-1}$] &
    Intensity [$10^{5}\,\rm{erg}\,\rm{cm}^{-2}\,\rm{s}^{-1}\,\rm{sr}^{-1}\,\AA^{-1}$]\\
    \midrule
    \endfirsthead
    \caption{continued.}\\
    \toprule
    Wavelength ($\Delta\lambda_{0}$) [$\AA$] & Intensity [$10^{-8}\,\rm{erg}\,\rm{cm}^{-2}\,\rm{s}^{-1}\,\rm{sr}^{-1}\,\rm{Hz}^{-1}$] &
    Intensity [$10^{5}\,\rm{erg}\,\rm{cm}^{-2}\,\rm{s}^{-1}\,\rm{sr}^{-1}\,\AA^{-1}$]\\
    \midrule
    \endhead
    \toprule
    \endfoot
     $-1.85$ & $  0.02$ & $  0.05$ \\
      $-1.81$ & $  0.03$ & $  0.05$ \\
      $-1.76$ & $  0.03$ & $  0.06$ \\
      $-1.72$ & $  0.03$ & $  0.06$ \\
      $-1.68$ & $  0.03$ & $  0.06$ \\
      $-1.63$ & $  0.03$ & $  0.07$ \\
      $-1.59$ & $  0.04$ & $  0.07$ \\
      $-1.55$ & $  0.04$ & $  0.08$ \\
      $-1.50$ & $  0.04$ & $  0.08$ \\
      $-1.46$ & $  0.04$ & $  0.09$ \\
      $-1.42$ & $  0.05$ & $  0.09$ \\
      $-1.38$ & $  0.05$ & $  0.10$ \\
      $-1.33$ & $  0.05$ & $  0.10$ \\
      $-1.29$ & $  0.05$ & $  0.11$ \\
      $-1.25$ & $  0.06$ & $  0.12$ \\
      $-1.20$ & $  0.07$ & $  0.14$ \\
      $-1.16$ & $  0.08$ & $  0.15$ \\
      $-1.12$ & $  0.08$ & $  0.17$ \\
      $-1.07$ & $  0.10$ & $  0.19$ \\
      $-1.03$ & $  0.11$ & $  0.22$ \\
      $-0.99$ & $  0.12$ & $  0.24$ \\
      $-0.94$ & $  0.13$ & $  0.27$ \\
      $-0.90$ & $  0.15$ & $  0.30$ \\
      $-0.86$ & $  0.17$ & $  0.34$ \\
      $-0.82$ & $  0.19$ & $  0.39$ \\
      $-0.77$ & $  0.22$ & $  0.45$ \\
      $-0.73$ & $  0.27$ & $  0.54$ \\
      $-0.69$ & $  0.33$ & $  0.67$ \\
      $-0.64$ & $  0.41$ & $  0.84$ \\
      $-0.60$ & $  0.52$ & $  1.05$ \\
      $-0.56$ & $  0.65$ & $  1.32$ \\
      $-0.51$ & $  0.84$ & $  1.70$ \\
      $-0.47$ & $  1.08$ & $  2.20$ \\
      $-0.43$ & $  1.41$ & $  2.87$ \\
      $-0.39$ & $  1.86$ & $  3.78$ \\
      $-0.34$ & $  2.46$ & $  5.00$ \\
      $-0.30$ & $  3.26$ & $  6.62$ \\
      $-0.26$ & $  4.14$ & $  8.42$ \\
      $-0.21$ & $  4.79$ & $  9.75$ \\
      $-0.17$ & $  4.85$ & $  9.85$ \\
      $-0.13$ & $  4.24$ & $  8.61$ \\
      $-0.08$ & $  3.57$ & $  7.27$ \\
      $-0.04$ & $  3.23$ & $  6.57$ \\
      $ 0.00$ & $  3.12$ & $  6.34$ \\
      $ 0.04$ & $  3.19$ & $  6.48$ \\
      $ 0.09$ & $  3.40$ & $  6.90$ \\
      $ 0.13$ & $  3.76$ & $  7.65$ \\
      $ 0.17$ & $  4.27$ & $  8.68$ \\
      $ 0.22$ & $  4.54$ & $  9.22$ \\
      $ 0.26$ & $  4.32$ & $  8.77$ \\
      $ 0.30$ & $  3.70$ & $  7.51$ \\
      $ 0.34$ & $  2.87$ & $  5.84$ \\
      $ 0.39$ & $  2.13$ & $  4.32$ \\
      $ 0.43$ & $  1.57$ & $  3.19$ \\
      $ 0.47$ & $  1.17$ & $  2.37$ \\
      $ 0.52$ & $  0.87$ & $  1.77$ \\
      $ 0.56$ & $  0.67$ & $  1.36$ \\
      $ 0.60$ & $  0.52$ & $  1.05$ \\
      $ 0.65$ & $  0.42$ & $  0.85$ \\
      $ 0.69$ & $  0.35$ & $  0.72$ \\
      $ 0.73$ & $  0.29$ & $  0.60$ \\
      $ 0.77$ & $  0.25$ & $  0.50$ \\
      $ 0.82$ & $  0.20$ & $  0.41$ \\
      $ 0.86$ & $  0.17$ & $  0.34$ \\
      $ 0.90$ & $  0.15$ & $  0.30$ \\
      $ 0.95$ & $  0.13$ & $  0.27$ \\
      $ 0.99$ & $  0.12$ & $  0.24$ \\
      $ 1.03$ & $  0.11$ & $  0.22$ \\
      $ 1.08$ & $  0.10$ & $  0.20$ \\
      $ 1.12$ & $  0.09$ & $  0.18$ \\
      $ 1.16$ & $  0.08$ & $  0.17$ \\
      $ 1.20$ & $  0.08$ & $  0.15$ \\
      $ 1.25$ & $  0.07$ & $  0.14$ \\
      $ 1.29$ & $  0.06$ & $  0.13$ \\
      $ 1.33$ & $  0.06$ & $  0.12$ \\
      $ 1.38$ & $  0.06$ & $  0.11$ \\
      $ 1.42$ & $  0.05$ & $  0.11$ \\
      $ 1.46$ & $  0.05$ & $  0.10$ \\
      $ 1.51$ & $  0.05$ & $  0.10$ \\
      $ 1.55$ & $  0.05$ & $  0.09$ \\
      $ 1.59$ & $  0.04$ & $  0.09$ \\
      $ 1.64$ & $  0.04$ & $  0.08$ \\
      $ 1.68$ & $  0.04$ & $  0.08$ \\
      $ 1.72$ & $  0.04$ & $  0.08$ \\
      $ 1.76$ & $  0.04$ & $  0.07$ \\
      $ 1.81$ & $  0.03$ & $  0.07$ \\
      $ 1.85$ & $  0.03$ & $  0.07$ \\
\end{longtable}

\begin{longtable}{ccc}
    \caption{Symmetrized reference Lyman-$\alpha$ profile
(Fig.~\ref{fig:reference}, right). The central wavelength
$\lambda_{0}$
    of the Lyman-$\alpha$ line is 1215.67\,$\AA$. The uncertainty of the
reference profile was estimated to be $\pm$(15/$\sqrt{3}$)\%.}\label{tab:App_symProf}\\
    \toprule
     Wavelength ($\Delta\lambda_{0}$) [$\AA$] & Intensity [$10^{-8}\,\rm{erg}\,\rm{cm}^{-2}\,\rm{s}^{-1}\,\rm{sr}^{-1}\,\rm{Hz}^{-1}$] &
    Intensity [$10^{5}\,\rm{erg}\,\rm{cm}^{-2}\,\rm{s}^{-1}\,\rm{sr}^{-1}\,\AA^{-1}$]\\
    \midrule
    \endfirsthead
    \caption{continued.}\\
    \toprule
    Wavelength ($\Delta\lambda_{0}$) [$\AA$] & Intensity [$10^{-8}\,\rm{erg}\,\rm{cm}^{-2}\,\rm{s}^{-1}\,\rm{sr}^{-1}\,\rm{Hz}^{-1}$] &
    Intensity [$10^{5}\,\rm{erg}\,\rm{cm}^{-2}\,\rm{s}^{-1}\,\rm{sr}^{-1}\,\AA^{-1}$]\\
    \midrule
    \endhead
    \toprule
    \endfoot
      $ 0.00$ & $  3.12$ & $  6.33$ \\
      $ 0.02$ & $  3.16$ & $  6.42$ \\
      $ 0.04$ & $  3.21$ & $  6.51$ \\
      $ 0.06$ & $  3.33$ & $  6.76$ \\
      $ 0.08$ & $  3.46$ & $  7.02$ \\
      $ 0.10$ & $  3.67$ & $  7.45$ \\
      $ 0.12$ & $  3.91$ & $  7.94$ \\
      $ 0.14$ & $  4.16$ & $  8.44$ \\
      $ 0.16$ & $  4.42$ & $  8.97$ \\
      $ 0.18$ & $  4.58$ & $  9.29$ \\
      $ 0.20$ & $  4.63$ & $  9.40$ \\
      $ 0.22$ & $  4.60$ & $  9.34$ \\
      $ 0.24$ & $  4.40$ & $  8.93$ \\
      $ 0.26$ & $  4.18$ & $  8.48$ \\
      $ 0.28$ & $  3.83$ & $  7.77$ \\
      $ 0.31$ & $  3.29$ & $  6.69$ \\
      $ 0.34$ & $  2.73$ & $  5.54$ \\
      $ 0.38$ & $  2.09$ & $  4.25$ \\
      $ 0.42$ & $  1.60$ & $  3.25$ \\
      $ 0.47$ & $  1.15$ & $  2.33$ \\
      $ 0.52$ & $  0.83$ & $  1.69$ \\
      $ 0.57$ & $  0.62$ & $  1.26$ \\
      $ 0.62$ & $  0.47$ & $  0.96$ \\
      $ 0.68$ & $  0.35$ & $  0.72$ \\
      $ 0.78$ & $  0.23$ & $  0.47$ \\
      $ 0.88$ & $  0.16$ & $  0.32$ \\
      $ 0.98$ & $  0.12$ & $  0.25$ \\
      $ 1.08$ & $  0.10$ & $  0.20$ \\
      $ 1.18$ & $  0.08$ & $  0.15$ \\
      $ 1.30$ & $  0.06$ & $  0.12$ \\
      $ 1.50$ & $  0.04$ & $  0.09$ \\
\end{longtable}
\renewcommand\arraystretch{1.0}
\newpage
\twocolumn

\onecolumn
\section{Coefficients describing the variation of solar
radiation in Lyman lines}\label{App:cycle}

\renewcommand\arraystretch{1.3}
\begin{longtable}{cccc}
    \caption{List of coefficients describing the variation of
    the solar radiation in the Lyman-$\alpha$ and higher Lyman
    lines. The coefficients were derived from a 400-day averaged
    LISIRD Lyman-$\alpha$ composite index (see
    Sect.~\ref{Sect:cycle_Lya}) for selected dates throughout the
    lifetime of SOHO. Coefficients for the Lyman-$\alpha$ line are
    computed with respect to the date of observations used in the
    present work (June 25, 2008). Coefficients for the higher Lyman
    lines are computed with respect to the date of observations
    used in \citet{1998ApJS..119..105W}, i.e. May 15, 1996.}\label{tab:App_coef}\\
    \toprule
    \multirow{2}{*}{Date} & Value of 400-day averaged & Coefficient for & Coefficient for \\
       & LISIRD index $[10^{-2}\,\mathrm{W/m}^2]$ & Lyman-$\alpha$ line & higher Lyman lines \\
    \midrule
    \endfirsthead
    \caption{continued.}\\
    \toprule
    \multirow{2}{*}{Date} & Value of 400-days averaged & Coefficient for & Coefficient for \\
       & LISIRD index $[10^{-2}\,\mathrm{W/m}^2]$ & Lyman-$\alpha$ line & higher Lyman lines \\
    \midrule
    \endhead
    \toprule
    \endfoot
   $1995/01/01$ & $0.644$ & $1.065$ & $1.037$ \\
   $1995/05/01$ & $0.642$ & $1.062$ & $1.033$ \\
   $1995/09/01$ & $0.635$ & $1.050$ & $1.022$ \\
   $1996/01/01$ & $0.627$ & $1.038$ & $1.010$ \\
   $1996/05/01$ & $0.622$ & $1.028$ & $1.001$ \\
   $\textbf{1996}/\textbf{05}/\textbf{15}$ & $0.621$ & $1.027$ & $1.000$ \\
   $1996/09/01$ & $0.620$ & $1.026$ & $0.998$ \\
   $1997/01/01$ & $0.622$ & $1.029$ & $1.001$ \\
   $1997/05/01$ & $0.634$ & $1.048$ & $1.020$ \\
   $1997/09/01$ & $0.654$ & $1.081$ & $1.052$ \\
   $1998/01/01$ & $0.684$ & $1.131$ & $1.101$ \\
   $1998/05/01$ & $0.725$ & $1.199$ & $1.168$ \\
   $1998/09/01$ & $0.765$ & $1.266$ & $1.232$ \\
   $1999/01/01$ & $0.798$ & $1.321$ & $1.286$ \\
   $1999/05/01$ & $0.822$ & $1.360$ & $1.324$ \\
   $1999/09/01$ & $0.847$ & $1.400$ & $1.363$ \\
   $2000/01/01$ & $0.876$ & $1.449$ & $1.410$ \\
   $2000/05/01$ & $0.895$ & $1.479$ & $1.440$ \\
   $2000/09/01$ & $0.894$ & $1.478$ & $1.439$ \\
   $2001/01/01$ & $0.879$ & $1.455$ & $1.416$ \\
   $2001/05/01$ & $0.874$ & $1.445$ & $1.407$ \\
   $2001/09/01$ & $0.894$ & $1.478$ & $1.439$ \\
   $2002/01/01$ & $0.905$ & $1.498$ & $1.458$ \\
   $2002/05/01$ & $0.908$ & $1.502$ & $1.463$ \\
   $2002/09/01$ & $0.864$ & $1.429$ & $1.392$ \\
   $2003/01/01$ & $0.823$ & $1.361$ & $1.325$ \\
   $2003/05/01$ & $0.797$ & $1.318$ & $1.283$ \\
   $2003/09/01$ & $0.768$ & $1.271$ & $1.237$ \\
   $2004/01/01$ & $0.755$ & $1.248$ & $1.215$ \\
   $2004/05/01$ & $0.734$ & $1.214$ & $1.182$ \\
   $2004/09/01$ & $0.715$ & $1.183$ & $1.151$ \\
   $2005/01/01$ & $0.702$ & $1.161$ & $1.130$ \\
   $2005/05/01$ & $0.687$ & $1.137$ & $1.107$ \\
   $2005/09/01$ & $0.668$ & $1.105$ & $1.076$ \\
   $2006/01/01$ & $0.657$ & $1.087$ & $1.058$ \\
   $2006/05/01$ & $0.648$ & $1.072$ & $1.043$ \\
   $2006/09/01$ & $0.643$ & $1.063$ & $1.035$ \\
   $2007/01/01$ & $0.636$ & $1.052$ & $1.024$ \\
   $2007/05/01$ & $0.626$ & $1.034$ & $1.007$ \\
   $2007/09/01$ & $0.615$ & $1.018$ & $0.991$ \\
   $2008/01/01$ & $0.609$ & $1.008$ & $0.981$ \\
   $2008/05/01$ & $0.605$ & $1.001$ & $0.974$ \\
   $\textbf{2008}/\textbf{06}/\textbf{25}$ & $0.605$ & $1.000$ & $0.973$ \\
   $2008/09/01$ & $0.603$ & $0.997$ & $0.970$ \\
   $2009/01/01$ & $0.602$ & $0.996$ & $0.970$ \\
   $2009/05/01$ & $0.604$ & $0.999$ & $0.973$ \\
   $2009/09/01$ & $0.613$ & $1.014$ & $0.987$ \\
   $2010/01/01$ & $0.625$ & $1.033$ & $1.006$ \\
   $2010/05/01$ & $0.640$ & $1.059$ & $1.031$ \\
   $2010/09/01$ & $0.653$ & $1.080$ & $1.051$ \\
   $2011/01/01$ & $0.674$ & $1.115$ & $1.086$ \\
   $2011/05/01$ & $0.704$ & $1.164$ & $1.133$ \\
   $2011/09/01$ & $0.738$ & $1.221$ & $1.189$ \\
   $2012/01/01$ & $0.752$ & $1.243$ & $1.210$ \\
   $2012/05/01$ & $0.761$ & $1.258$ & $1.225$ \\
   $2012/09/01$ & $0.749$ & $1.238$ & $1.205$ \\
   $2013/01/01$ & $0.753$ & $1.245$ & $1.212$ \\
   $2013/05/01$ & $0.757$ & $1.253$ & $1.220$ \\
   $2013/09/01$ & $0.777$ & $1.285$ & $1.251$ \\
   $2014/01/01$ & $0.793$ & $1.311$ & $1.276$ \\
   $2014/05/01$ & $0.806$ & $1.333$ & $1.298$ \\
   $2014/09/01$ & $0.813$ & $1.344$ & $1.308$ \\
   $2015/01/01$ & $0.808$ & $1.337$ & $1.301$ \\
   $2015/05/01$ & $0.791$ & $1.308$ & $1.274$ \\
   $2015/09/01$ & $0.762$ & $1.261$ & $1.227$ \\
   $2016/01/01$ & $0.727$ & $1.202$ & $1.170$ \\
   $2016/05/01$ & $0.703$ & $1.162$ & $1.131$ \\
   $2016/09/01$ & $0.671$ & $1.109$ & $1.080$ \\
   $2017/01/01$ & $0.651$ & $1.076$ & $1.047$ \\
   $2017/05/01$ & $0.637$ & $1.054$ & $1.026$ \\
   $2017/09/01$ & $0.629$ & $1.040$ & $1.012$ \\
   $2018/01/01$ & $0.621$ & $1.028$ & $1.001$ \\
   $2018/05/01$ & $0.612$ & $1.013$ & $0.986$ \\
   $2018/09/01$ & $0.606$ & $1.002$ & $0.975$ \\
   $2019/01/01$ & $0.604$ & $0.999$ & $0.972$ \\
   $2019/05/01$ & $0.599$ & $0.991$ & $0.965$ \\
   $2019/09/01$ & $0.599$ & $0.991$ & $0.965$ \\
   $2020/01/01$ & $0.599$ & $0.990$ & $0.964$ \\
   $2020/05/01$ & $0.601$ & $0.993$ & $0.967$ \\
\end{longtable}

\end{appendix}

\end{document}